\documentclass{article}

\usepackage[nonatbib,final]{nips_2018}

\usepackage[utf8]{inputenc} 
\usepackage[T1]{fontenc}    
\usepackage{hyperref}       
\usepackage{url}            
\usepackage{booktabs}       
\usepackage{amsfonts}       
\usepackage{nicefrac}       
\usepackage{microtype}      
\usepackage{graphicx}
\usepackage{amsmath}
\usepackage{gensymb}
\usepackage{color}
\usepackage{pgfplots}
\usepackage{subcaption}
\usepackage{float}
\usepackage{caption}
\usepackage{subcaption}
\usepackage{graphicx}
\usepackage{booktabs}

\usepackage{abstract}

\title{Efficient allocation of law enforcement resources using predictive police patrolling\vspace{-0.05in}}

\author{
  Mateo Dulce \\
  University of los Andes and Quantil\\
  \texttt{mate.dulce@quantil.com.co} \\
  \And
  Simon Ramirez \\
  University of los Andes and Quantil\\
  \texttt{simon.ramirez@quantil.com.co} \\
  \And
  Alvaro Riascos \\
  University of los Andes and Quantil\\
  \texttt{ariascos@uniandes.edu.co} 
}

\begin{document}
\vspace{-0.25in}
\maketitle
\vspace{-0.25in}
\begin{abstract}

Efficient allocation of scarce law enforcement resources is a  hard problem to tackle. In a previous study (forthcoming Barreras et.al (2019)) it has been shown that a simplified version of the self-exciting point process explained in Mohler et.al (2011), performs better predicting crime in the city of Bogot\'{a} - Colombia, than other standard hotspot models such as plain KDE or ellipses models. This paper fully implements the Mohler et.al (2011) model in the city of Bogot\'{a} and explains its technological deployment for the city as a tool for the efficient allocation of police resources. 
  
\end{abstract}

\section*{Introduction}
\vspace{-0.1in}
Criminality is one of the biggest challenges mega-cities face. Among many other decisions, policy makers have to efficiently allocate scarce law enforcement resources on a vast and highly dynamic environment. This is a hard problem with no trivial solution. For example, during 2012 and 2015 all murders and 25\% of all crimes in Bogota took place in just 2\% of street segments. Yet, these same road segments received less than 10\% of effective police patrolling time. Understanding the spatial and temporal dynamics of these so-called \textit{hotspots} is needed to make highly effective police patrolling possible. In this paper we develop a \textit{self exciting point process} model to predict crime and present partial results of its deployment on field scenarios in Bogotá, Colombia. 

This paper is organized as follows. Section 2 describes the theoretical model used to approach crime prediction. Section 3 describes the training of the model and section 4 its validation. Finally, section 5 presents the technological deployment and visualization of model. 

\section*{The model}

The model developed to predict crime occurrences in Bogotá, Colombia, follows closely the methodology proposed by Mohler et al. (2011) in their work \textit{Self-Exciting Point Process Modeling of Crime} (Mohler et.al, 2011). This model is constructed under three assumptions:
\begin{enumerate}
    \item Criminality concentrates in specific areas of the city. 
    \item Higher incidence of crime at certain times of the day and certain days of the week.
    \item Crime spread from one place to another like a disease.  
\end{enumerate}

With this in mind, crimes are classified between background and aftershock events, the former being those that arise independently given their spatio-temporal location, while the latter occur as triggering of past crimes nearby. Crime appearance is modeled as a self-exciting point process in which the past occurrence of crimes increases the probability of new crimes occurring in the future.

A spatio-temporal point process $N(x,y,t)$ is uniquely characterized by its conditional intensity $\lambda(x,y,t)$, which can be defined as the expected number of points falling in an arbitrarily small spatio-temporal region, given the points history $\mathcal{H}_t$ occurred until $t$:

\begin{equation}
\label{defIntensity}
\lambda(x,y,t) = \lim_{\Delta x, \Delta y, \Delta t \downarrow 0} \frac{E[N\{(x,x+\Delta x)\times (y,y+\Delta y)\times (t,t+\Delta t)\}|\mathcal{H}_{t}]}{\Delta x \Delta y \Delta t}.
\end{equation}

For the purpose of crime prediction and according to the initial assumptions on the behavior of crime occurrence, it is assumed that the conditional intensity takes the following functional form

\begin{equation}
\label{intensity}
\lambda(x,y,t) = \mu(x,y)\nu(t) + \sum_{\{k:t_k<t\}}g(x-x_{k}, y-y_{k}, t-t_{k}),
\end{equation}
where $\mu(x,y)$ and $\nu(t)$ captures background crimes appearance patterns according to their spatial and temporal location, respectively. In a similar fashion, $g(x-x_{k}, y-y_{k}, t-t_{k})$ captures how crime $(x_{k}, y_{k}, t_{k})$ propagates to other spatio-temporal locations.

\section*{Training}

We worked with criminal data from the \textit{Delinquential, Contraventional and Operative Statistical Information System} (SIEDCO) from the National Police of Colombia. This dataset contains georeferenced crimes occurred in Bogotá during 2017, along with the day and time of the crime. We aggregate the data according to the patrol shifts of Bogotá police department for 3 daily and 21 weekly shifts. Then we construct the \textit{circular\_time} variable that summarizes the day and time of the week in which a crime occurs, and \textit{linear\_time} that keeps the temporary record of the occurred crimes.

\textit{Circular\_time} is the input variable of the function $\nu$ which looks for temporal patterns of crime occurrence, while \textit{linear\_time} is used in the triggering function $g$ to study the temporal distance between the occurred crimes. Finally, the function $\mu$ use the latitude and longitude coordinates of historic crimes to find spatial patterns. 

To estimate the conditional intensity function, is necessary to differentiate between background crimes and those triggered by past crimes, and use each of these families of data to estimate the functions: $\mu$ and $\nu$ with background events and $g$ with aftershock crimes. The training of the model is then based on \textit{stochastic declustering} techniques and Kernel density estimation.

Assuming that the functional form of the conditional intensity is correct, the probability that crime $i$ was triggered by crime $j$ is:
\begin{equation}
\label{probRep}
p_{ij} = \frac{g(x_{i}-x_{j}, y_{i}-y_{j}, t_{i}-t_{j})}{\lambda(x_{i},y_{i},t_{i})}.
\end{equation}
On the other hand, the probability that crime $i$ is a background event is given by:
\begin{equation}
\label{probTrans}
p_{ii} = \frac{\mu(x_{i}, y_{i})\nu(t_{i})}{\lambda(x_{i},y_{i},t_{i})}.
\end{equation}

Let $P$ denote the matrix with entries $p_{ij}$. Note that $P$ is an upper triangular matrix given that a crime cannot be triggered by a future event, and that, by definition of $\lambda(\cdot)$, columns sum to one. Then we perform an iterative algorithm until the matrix $P$ converges in the following way:

\begin{enumerate}
    \item $P_{0}$ is initialized assuming that the crime triggering process decays exponentially in time and behaves as a bivariate normal distribution on the spatial coordinates (Rosser and Cheng, 2016):
    $$p_{ij} = exp(-\alpha(t_{i}-t_{j}))exp\left(\frac{-(x_{i}-x_{j})^{2}-(y_{i}-y_{j})^{2}}{2\beta^{2}}\right), ~~~ i\leq j, $$
    normalizing its columns such that each one sum to one. In all the exercises performed we found that the parameters $\alpha = 0.03$ y $\beta=100$ yield to consistent results. 
    \item Given $P_{n-1}$, sample background events $\{(x_{i}^{b},y_{i}^{b},t_{i}^{b})\}_{i=1}^{N_{b}}$ and triggered crimes $\{(\Delta x_{i}^{r},\Delta y_{i}^{r},\Delta t_{i}^{r})\}_{i=1}^{N_{r}}$ where $(\Delta x_{i}^{r}, \Delta y_{i}^{r}, \Delta t_{i}^{r})$ denotes the spatio-temporal distance of crime $i$ to the triggering crime.
    \item Estimate functions $\mu_{n}$ and $\nu_
    {n}$ using the sampled background crimes, and $g_{n}$ with the sampled triggered crimes, using Kernel density estimation. 
    \item Update matrix $P_{n}$ using estimated functions $\mu_{n}$, $\nu_{n}$ and $g_{n}$ in the previous step, and relations (\ref{probRep}) and (\ref{probTrans}). If $||P_{n}-P_{n-1}||_{2} \geq \epsilon$, go to step 2\footnote{$||.||_{2}$ denotes matrix norm in $L_{2}$. $\epsilon=0.01$ was used as convergence parameter of $P$.}.
\end{enumerate}

Note that in each iteration the number of background and aftershock events varies, but the total number of crimes remains constant, $N^{b}+N^{a}=N$. Thus, in the Kernel density estimation, we use a variable bandwidth that updates in each iteration using maximum likelihood with the sampled events for each function. 

Finally, we obtain functions $\mu$, $\nu$ and $g$ from the training process and construct the conditional intensity $\lambda$ with them. Then, to predict the occurrence of crimes, we evaluate the intensity function in the spatial coordinates and shifts of interest. For this, we need to specify these two dimensions for estimation and to make it useful for police authorities, according to their operating manners.

\begin{enumerate}
\item Time dimension: eight hours slots.
\item Identify two main hotspots per locality.
\end{enumerate}

\begin{figure}[H]
    \centering
    \begin{subfigure}[t]{0.45\textwidth}
        \centering
        \includegraphics[width=\linewidth]{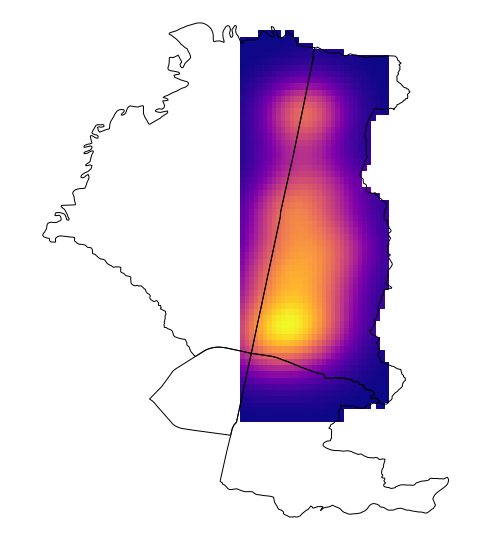} 
        \caption{Generic} \label{fig:timing1}
    \end{subfigure}
    \hfill
    \begin{subfigure}[t]{0.45\textwidth}
        \centering
        \includegraphics[width=\linewidth]{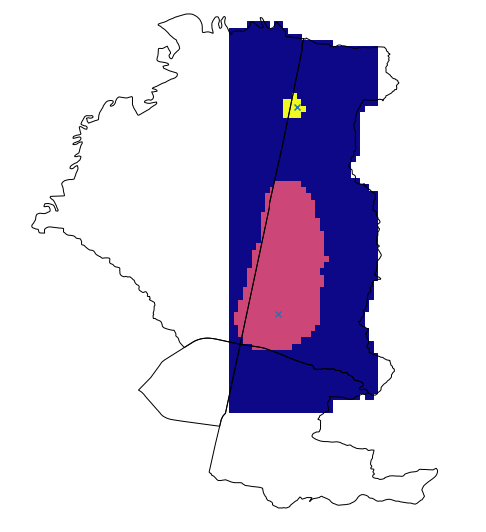} 
        \caption{Competitors} \label{fig:timing2}
    \end{subfigure}

\end{figure}

\section*{Validation}

To evaluate the predictive capacity of the model of crime as a self-exciting point process, and to select the right parameters that maximizes such predictive capacity, we use the standard measure of \textit{Hit Rate} that indicates the portion of crimes correctly predicted by the model. For this, we divide the city in uniform cells and compute the intensity of each cell using a Monte Carlo method. Then, we choose the critical cells (hotspots) and evaluate how many of the known crimes occurred in these cells. 

$$\textnormal{Hit Rate} = \frac{\textnormal{\# Crimes in the predicted hotspots}}{\textnormal{Total \# crimes}}.$$

The training and validation process were performed on the 10\% of the city cells that include Santa Fe sector. We train the proposed model with criminal data in this area during ten weeks (from June 22, 2017 to August 31, 2017), and tests its predictive accuracy checking the crimes occurred in the following four weeks (from September 1 to 28, 2017). The validation process shows that the model of crime as a self-exciting point process trained using variable bandwidth predicts a greater number of crimes, on average, than the model with fixed bandwidth or a plain KDE. 

\begin{itemize}
\item Accuracy per unit of area identified:
\begin{align*}
\text{PAI}= \dfrac{\text{Hit Rate}}{\text{Percentage of Area}}\\
\vspace{40pt}
\text{Hit Rate}= \dfrac{\text{Crimes predicted in Hotspots}}{\text{Total Crimes}}\\
\vspace{40pt}
\text{Percentage Area}= \dfrac{\text{Area of Hotspots}}{\text{Total Area}}
\end{align*}

\item Hit Rate with 7 weeks of training data and 10\% of covered area (i.e., hotspots):
\begin{center}
 \begin{tabular}{c c c c c}
\toprule
 Prediction &&  KDE & fixed bw & variable bw  \\ [0.5ex] 
 \midrule
 Week 1 && 0,42 & 0,44 & 0,57 \\ 
 Week 2 && 0,44 & 0,46 & 0,59 \\
 Week 3 && 0,53 & 0,54 & 0,62 \\
 \midrule \midrule
 Average && 0,46 & 0,48 & 0,59 \\ [1ex] 
 \bottomrule
\end{tabular}
\end{center}
\end{itemize}

Finally, to assess whether the predictive accuracy of the models differ statistically, we used the non-parametric \textit{Wilcoxon signed-rank test} to compare the obtained samples of crime prediction. The results show that the self-exciting point process modeling of crime performs statistically better predicting crime in the city of Bogotá, Colombia, than other state-of-the-art crime prediction models. 

\begin{center}
 \begin{tabular}{l c c}
 \toprule
 \multicolumn{1}{c}{Model} && p-value \\ [0.5ex] 
 \midrule
 fixed bw vs. KDE && 0,061\\ 
 fixed bw vs. varible bw && 0,030\\
 variable bw vs. KDE && 0,016\\ 
 \bottomrule
\end{tabular}
\end{center}

\section*{Field deployment and visualization}
We jointly developed a hybrid (web and mobile) application with local law enforcement authorities\footnote{Secretaría de Seguridad, Convivencia y Justicia de Bogotá (SSCJ) and Policía Metropolitana de Bogotá (MEBOG)} and Colombia's main research center on security studies\footnote{Centro de Estudios sobre Seguridad y Drogas (CESED) - Universidad de los Andes} to deploy our model in real-life field scenarios in Bogota. Our app is available to both planning agents via web browsers and to field agents via device specific native containers. The application is integrated to local law enforcement information systems and uses the most recent available crime data to calibrate self exciting process models. The app offers users two main features:

\begin{itemize}
    \item Layout a crime intensity heatmap over neighborhoods under agent surveillance for weekly schedules. (see Figure 2.a).
    
    \item Display critical hotspots in neigborhoods under agent surveillance for weekly schedules. (see Figure 2.b).
\end{itemize}

The application is under pilot trial in 10 crime ridden neigborhoods in Bogota since November 2017. Early results are encouraging. Police agents have appropriated the app as part of their decision making process and are looking forward to further development. A more extensive and rigorous randomized control trial to account for causal effects on safety and liveability related variables on 500 neighborhoods is currently being developed.

\begin{figure}
    \centering
    \includegraphics[width=0.9\linewidth]{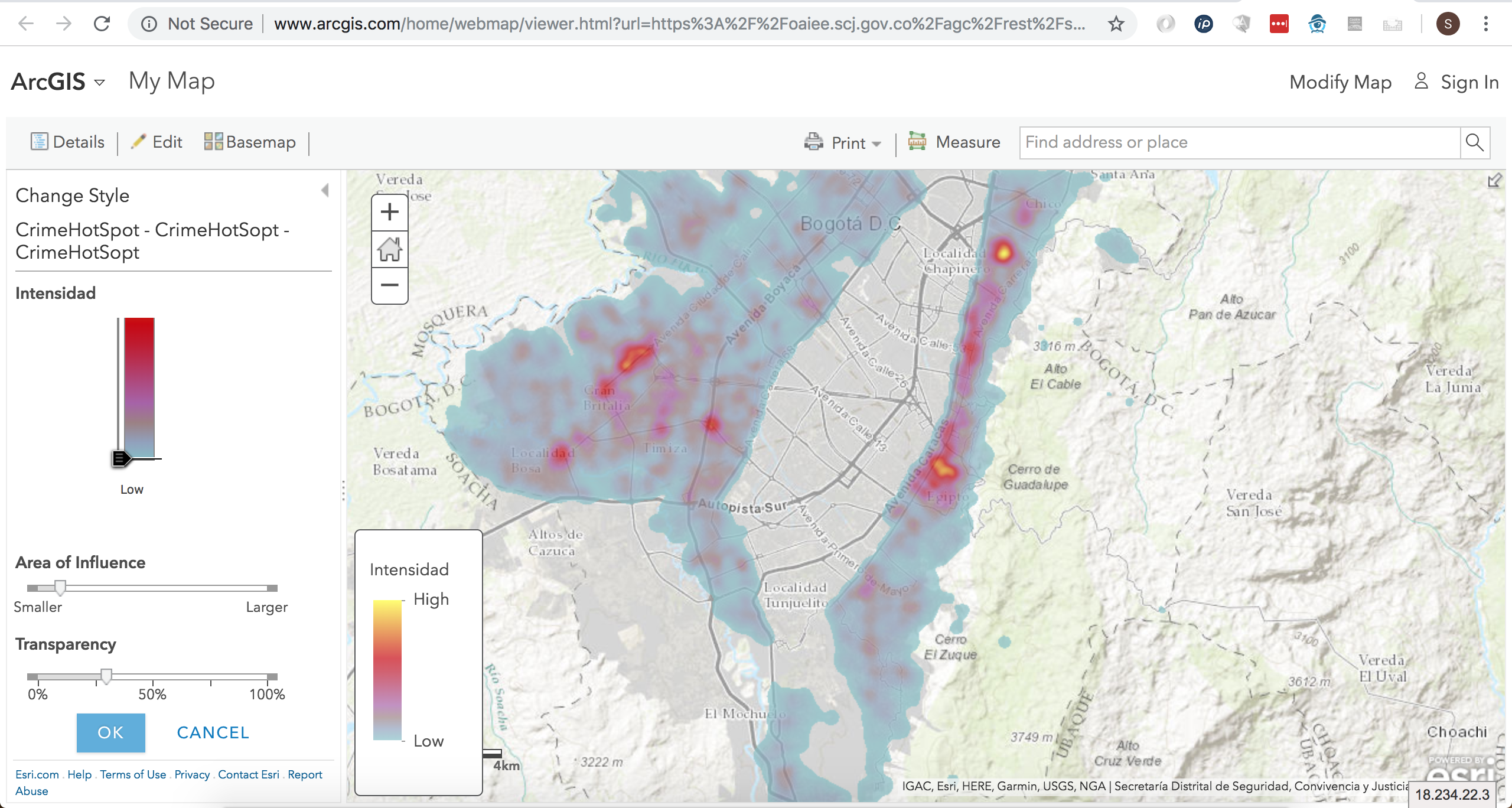}
    \caption{Web based view of crime intensity heatmaps over Bogotá}
\end{figure}

\begin{figure}
    \centering
    \includegraphics[width=0.9\linewidth]{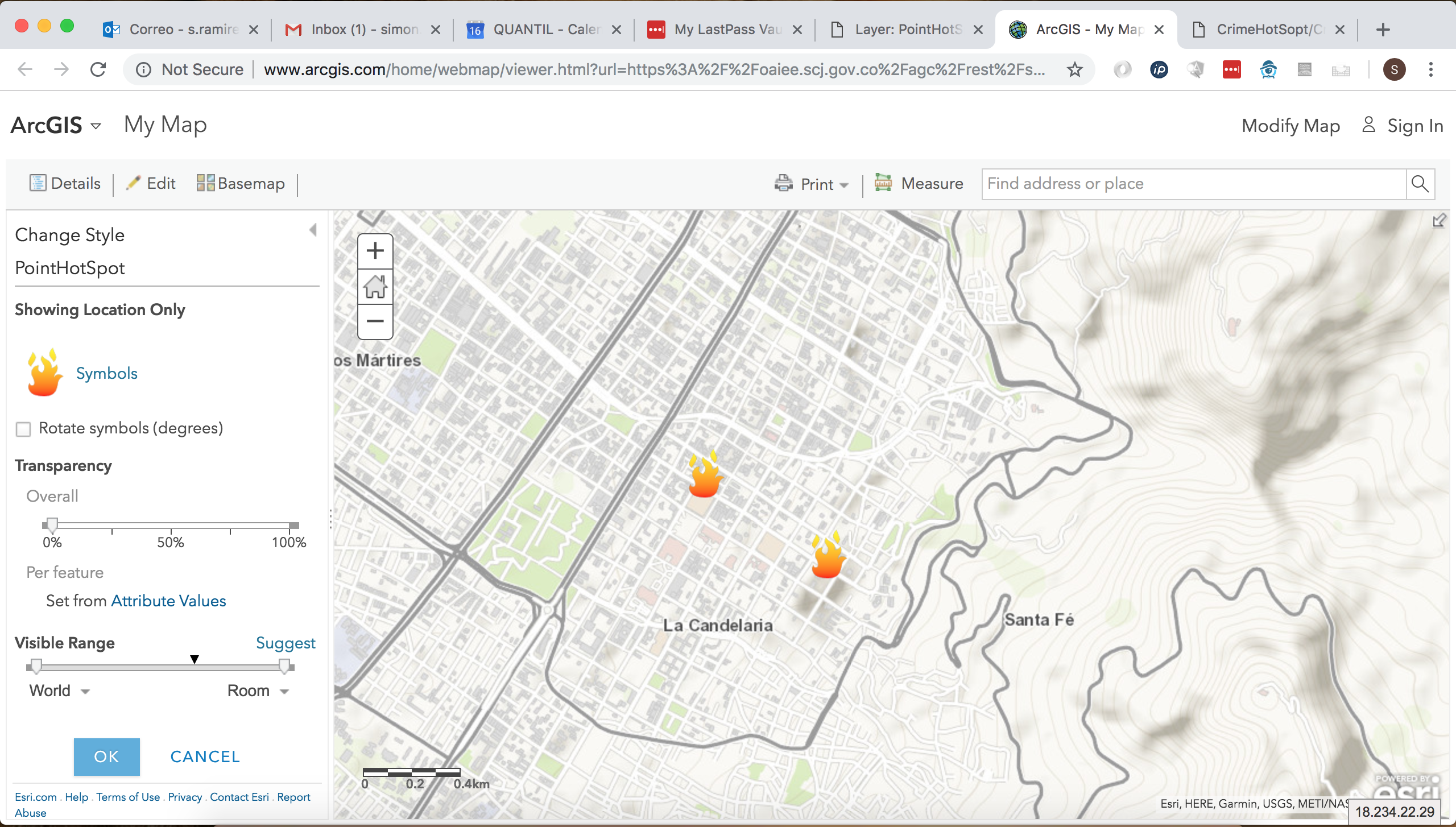}
    \caption{Web based view of critical hotspots over Bogotá}
\end{figure}

\nocite{*} 
\bibliographystyle{plain} 

\end{document}